\documentclass[a4paper]{article}
\usepackage{amsmath,graphicx}
\usepackage{amsmath,amssymb,amsfonts}
\usepackage{algorithmic}
\usepackage{graphicx}
\usepackage{textcomp}

\usepackage{mathtools}
\usepackage{fixmath}
\usepackage[normalem]{ulem}
\usepackage{multirow}
\usepackage[dvipsnames,table,xcdraw]{xcolor}
\usepackage{hyperref}
\usepackage{cite}

\usepackage{INTERSPEECH2022}

\title{Emphasis control for parallel neural TTS}
\name{Shreyas Seshadri, Tuomo Raitio, Dan Castellani, Jiangchuan Li}
\address{Apple}
\email{\vspace{-4mm}}

\begin{document}

\maketitle
\begin{abstract}
Recent parallel neural text-to-speech (TTS) synthesis methods are able to generate speech with high fidelity while maintaining high performance. However, these systems often lack control over the output prosody, thus restricting the semantic information conveyable for a given text. This paper proposes a hierarchical parallel neural TTS system for prosodic emphasis control by learning a latent space that directly corresponds to a change in emphasis. Three candidate features for the latent space are compared: 1) Variance of pitch and duration within words in a sentence, 2) Wavelet-based feature computed from pitch, energy, and duration, and 3) Learned combination of the two aforementioned approaches. At inference time, word-level prosodic emphasis is achieved by increasing the feature values of the latent space for the given words. Experiments show that all the proposed methods are able to achieve the perception of increased emphasis with little loss in overall quality. Moreover, emphasized utterances were preferred in a pairwise comparison test over the non-emphasized utterances, indicating promise for real-world applications.
\end{abstract}

\section{Introduction}

State-of-the-art neural text-to-speech (TTS) synthesis methods are able to generate speech with high fidelity while maintaining high performance \cite{ren2019fastspeech, ren2020fastspeech, donahue2020end}. However, the input to the TTS system does not often contain enough information in order to reconstruct the desired speech signal with the intended meaning and nuances. Therefore, TTS systems often generate average prosody learned from the database. This is disadvantageous when there is a need to convey a specific style or emotion, or highlight the part of the sentence that contributes new, non-derivable, or contrastive information.

A prosodically highlighted region of a speech signal is said to be \textit{emphasised}, \textit{prominent}, or in \textit{lexical focus}. This involves complex variations of many aspects of the speech signal, such as pitch, phoneme duration, and spectral energy \cite{eriksson2013acoustics, eriksson2015acoustics, eriksson2016acoustics, barbosa2013robustness}. Moreover, the combination of these acoustic feature variations that contribute to perceived emphasis is language dependent \cite{Eriksson2018, Eriksson2020}. Therefore, it is not feasible to manually adjust speech features to create a natural sounding perception of emphasis. This paper focuses on providing emphasis control in parallel neural TTS based on a learned latent space so that any part of the utterance may be emphasized by a desired degree.

Several methods have been explored within neural TTS to add controllability to the prosody of synthetic speech. For instance, with prosody transfer \cite{klimkov2019fine}, a separate prosody encoder is learned that encodes the prosodic information in speech that is not present in the text input. During inference, a reference speech sample with the appropriate prosody can be used as an input along with the text to ``copy" the prosody from the reference to the output synthetic speech. Such approaches can provide a high degree of control, however, a reference speech sample must be recorded for each utterance with the intended prosody, which is not always practical in real applications.

Some approaches learn an unsupervised latent space that encodes the prosodic information in the speech signal that is not described by the text input \cite{skerry2018towards}. This latent space can be represented as a learned set of style tokens \cite{wang2018style} or constrained to be a Gaussian (mixture) using a variational probabilistic approach \cite{hsu2017learning, akuzawa2018expressive, karanasou2021learned, hsu2018hierarchical}. Different regions of the learned latent space encode different aspects of prosody. During inference, a prosody encoding from a specific region of the latent space can be sampled to obtain synthesis with appropriate prosody. However, this approach requires a thorough analysis of the learned latent space, and manually identifying the mapping from the latent space to a specific prosody or style, which can be challenging for high dimensional latent spaces. Moreover, as the latent space is learned without supervision, it represents all of the acoustic differences in the speech signal that are not represented by the corresponding text, possibly including undesired variation (such as background noise or the effect of audio channel) and entangled representations. This makes it difficult to use such a mapping to achieve specific and predictable changes in prosody, further adding to the difficulty of using these methods in real-world applications.

A more interpretable latent space of prosody can be achieved by using well-known acoustic speech features, such as pitch, energy, phoneme duration, and spectral tilt \cite{Shechtman_2019, raitio2020controllable, raitio2022prosody}. For example, the methods in \cite{raitio2020controllable, raitio2022prosody} allow for modifications in prosody such as speaking rate, pitch, pitch range, or speaking style to softer or louder, and the method in \cite{raitio2022prosody} has been shown to provide word-wise emphasis control. However, a learned representation dedicated directly to emphasis may be useful.

Prosodic emphasis control in TTS has been explored in several previous studies relating to autoregressive attention-based models. The method in \cite{Yasuda2019} used accent-type sequence along with the input phoneme sequence in a Tacotron 2-based \cite{shen2018natural} system to handle Japanese accentual-type labels. The method in \cite{suni2020prosodic} extended the DCTTS \cite{tachibana2018efficiently} system to enable prosodic control using a set of features calculated using the continuous wavelet transform (CWT) on a combination of pitch, energy, and word duration signals \cite{suni2017hierarchical}. The method in \cite{shechtman2021supervised} modified the Tacotron 2 system to use variance-based features of pitch and phoneme duration. Here the text is used to predict the latent space of these features, which can then be modified as needed during inference for emphasis control. The study in \cite{liu2021controllable} introduces a forward attention-based TTS system that is capable of emphasis control.
While these approaches are able to achieve a certain degree of emphasis control, attention-based autoregressive TTS systems, in general, have certain downsides. These include inefficiencies at both at training and inference with modern parallel computing hardware and instabilities at inference time due to the attention-based alignment, resulting in skipping or repetition of words and mumbling.

This paper proposes a parallel neural TTS model based on FastSpeech 2 \cite{ren2020fastspeech} that is capable of emphasis control\footnote{Speech samples can be found at \url{https://apple.github.io/parallel-tts-emphasis-control/}.} using features similar to \cite{shechtman2021supervised} and \cite{suni2020prosodic}. These higher-level features are first predicted from text, and are later used to predict the phoneme-wise pitch, energy, and duration features in a hierarchical structure similar to \cite{malisz19_ssw, raitio2022prosody}. This ensures that the emphasis features directly influence only the prediction of the pitch, energy, and phoneme duration, while retaining the quality of output synthetic speech. Additionally, a constrained 1D combination of the emphasis features \cite{shechtman2021supervised, suni2020prosodic} is also explored, such that the intuitive relationship between the learned latent space and prosodic emphasis is retained.

\section{Emphasis control model}
\label{sec:emph}

\begin{figure}[tb]
    \centering
    \includegraphics[width=0.9\linewidth]{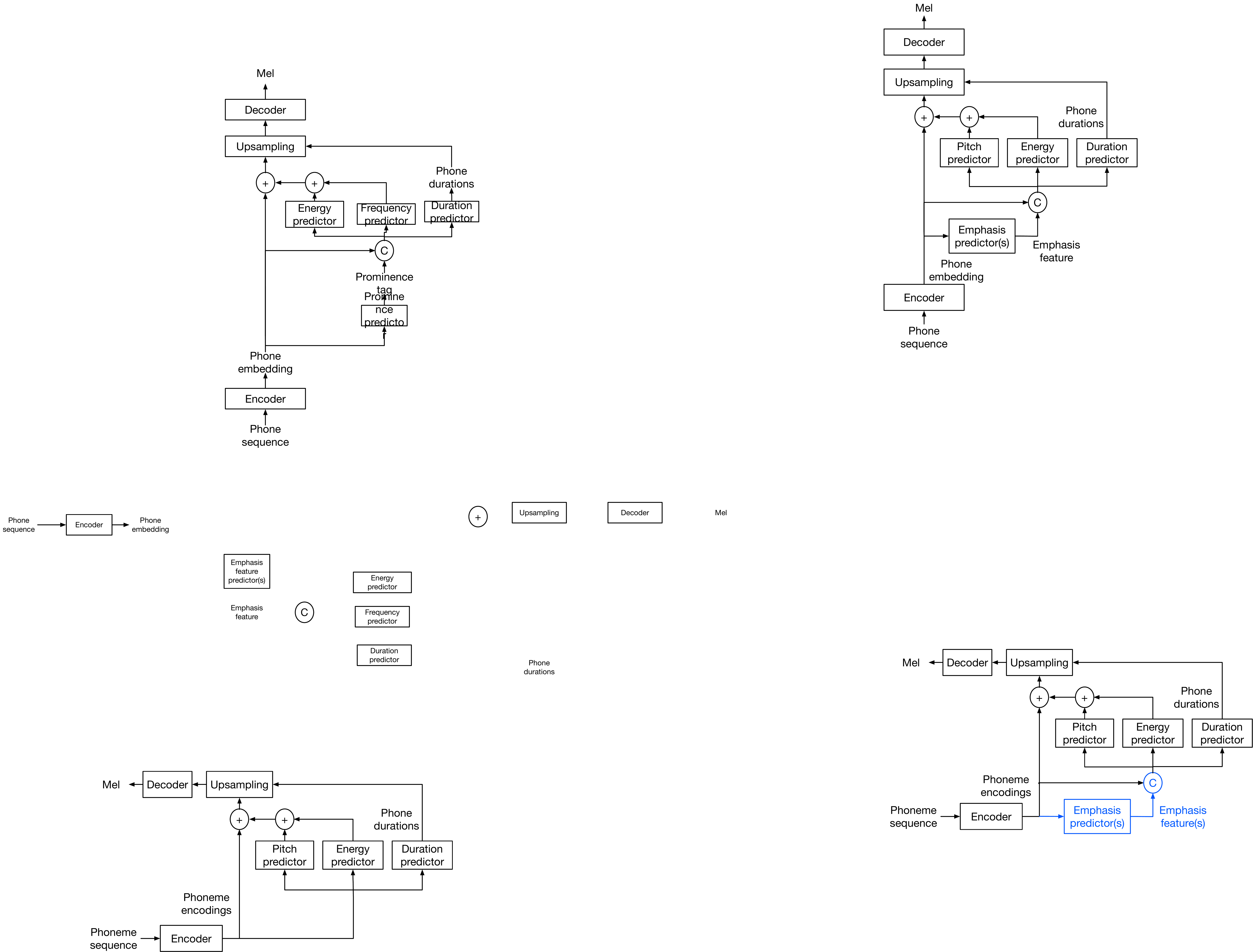}
    \vspace{-1.5mm}
    \caption{Architecture of the proposed TTS model. The components of the baseline FastSeepch 2-based system are shown in black while the components contributing to the proposed hierarchical emphasis control are shown in blue.}
    \vspace{-4mm}
    \label{fig:block_small}
\end{figure} 

The baseline model architecture is based on the FastSpeech 2 \cite{ren2020fastspeech} parallel neural TTS model that converts an input phoneme sequence to a Mel spectrogram. The model consists of an encoder that converts the phoneme sequence to phoneme encodings. As in \cite{ren2020fastspeech}, the encoder consists of an embedding layer that converts the phoneme sequence to phoneme embeddings. Next, a series of feed-forward transformer (FFT) \cite{ren2019fastspeech} blocks take in the phoneme embeddings with positional encodings, and output the phoneme encodings. Each FFT block consists of a self-attention layer \cite{vaswani2017attention} and 1-D convolution layers \cite{gehring2017convolutional} along with layer normalization \cite{ba2016layer} and dropout \cite{srivastava2014dropout}. The phoneme encodings are then fed to the  feature predictors that predict the phoneme-wise pitch, duration and energy. The feature predictors consist of 1-D convolution layers, layer normalization and dropout similar to the variance adaptors used in \cite{ren2020fastspeech}. The predicted phoneme-wise pitch and energy features are quantized, passed though an embedding layer, and added to the phoneme encodings, to form the decoder input. All the feature predictors are trained using teacher forcing. The decoder inputs are upsampled based on the predicted phoneme-wise durations. The FastSpeech 2 \cite{ren2020fastspeech} decoder which consists of FFT blocks is replaced here with a series of dilated convolution stacks, which improves model inference speed. This model achieves 150$\times$ faster than real-time synthesis on a GPU and 100$\times$ on a mobile device.

In order to enable emphasis control, the baseline architecture is extended to include a phoneme-wise emphasis feature predictor, which uses the same architecture as in the pitch, energy and duration feature predictors. The emphasis predictor takes the phoneme encodings as inputs and output the emphasis features. The emphasis features are broadcast concatenated with the phoneme encodings before being fed to the pitch, energy and duration feature predictors. This hierarchical structure allows for interpretability and fine-grained control. The model architecture is shown in Fig.~\ref{fig:block_small}.

\subsection{Emphasis features}
\label{sec:emph_feat}

This study explores three sets of emphasis features that enable easy and interpretable emphasis control. For each of these features, a separate parallel neural TTS model is trained and compared in Sec.~\ref{sec:obj} and \ref{sec:subj}. Each of these features is calculated at the word-level and are normalized as in \cite{shechtman2021supervised} using the linear mapping of $[-3\sigma^2, 3\sigma^2] \rightarrow [-1,1]$, where $\sigma^2$ is the variance of each feature in the training data.

\subsubsection{Variance-based features (VarianceBased)}
\label{sec:var}

Variance-based features consist of two features, word-wise pitch and phoneme duration variances \cite{shechtman2021supervised}:

\vspace{-2.5mm}
\begin{equation}
\label{eq:var_feats}
\begin{aligned}
    \text{Pitch variance} &= W_{F0} - S_{F0}, \\
    \text{Phoneme duration variance} &= W_{dur} - S_{dur},
\end{aligned}
\end{equation}
where $W_{dur}$ and $S_{dur}$ are the average per-phoneme durations along words and sentences respectively, and $W_{F0}$ and $S_{F0}$ are the log-pitch spreads along words and sentences respectively. The log-pitch spread is defined as the difference between the 95$^{th}$ and 5$^{th}$ percentiles of log-pitch. Unlike in \cite{shechtman2021supervised}, we did not include $S_{F0}$ and $S_{dur}$ to the list of trained features as they did not provide any benefits in our study on emphasis control. The pitch and phoneme duration variances are modeled by two separate emphasis predictors in Fig.~\ref{fig:block_small}.

\subsubsection{Wavelet-based prosody feature (WaveletBased)}
\label{sec:prom}

The wavelet-based prosody feature is based on the continuous wavelet transform (CWT) of a weighted sum of the pitch, energy, and duration signals (calculated as a series of deltas at the level of a phoneme, syllable, word or their combination) of speech as in \cite{suni2017hierarchical, suni2020prosodic}. The CWT decomposition of this 1-D signal results in scales that can be associated with levels of prosodic hierarchy. A reduced tree representation of the scaleogram is derived using a method called lines of maximum amplitude \cite{suni2017hierarchical} that connects regions of local maxima. Integrating the scale values of these local maxima at the word-level yields the continuous wavelet prosody feature. The \textit{WaveletBased} feature is modeled by a single emphasis predictor.

\subsubsection{Combined features (Combined)}

The \textit{VarianceBased} and \textit{WaveletBased} features are based on different methods and thus they may have a different relation to the actual emphasis of speech. The \textit{Combined} approach attempts learn a combination of these features. Three separate emphasis predictors are used to for the two \textit{VarianceBased} features and \textit{WaveletBased} feature, after which an unbiased linear mapping layer is used to learn a weighted combination of the three features. The weights of the linear layer are constrained to be positive to ensure that increasing values of the \textit{Combined} feature correspond to increasing emphasis. They are also constrained to sum to one, which ensures that the range of the feature values is the same as the \textit{VarianceBased} and \textit{WaveletBased} features.

\section{Experiments}
\label{sec:expt}

\subsection{Data}
\label{ssec:data}

We use two American English voices, a high-pitched voice with 36 hours of audio (Voice 1), and a low-pitched voice with 23 hours of audio (Voice 2). All speech data were sampled at 24~kHz. For testing, we curated 192 sentences with a single highlighted word to be emphasized per sentence.

\subsection{System specifications}

The input of the emphasis control model is a phoneme sequence with punctuation and word boundaries. The encoder of the model, shown in Fig.~\ref{fig:block_small}, has 4 FFT layers, each with a self-attention layer having 2 attention heads and 256 hidden units, and two 1-D convolution layers, each having a kernel size of 9 and 1024 filters. The decoder has 2 dilated convolution blocks with six 1-D convolution layers with dilation rates of 1, 2, 4, 8, 16, and 32, respectively, kernel size of 3, and 256 filters. The feature predictors have two 1-D convolution layers, each having a kernel size of 3 and 256 filters. The dropout rate was set to 0.2 and $\epsilon$ for the layer normalization was set to $10^{-6}$.

The \textit{WaveletBased} features were calculated using the open source Wavelet Prosody Toolkit\footnote{\url{https://github.com/asuni/wavelet_prosody_toolkit}} \cite{suni2017hierarchical}. The hyper-parameters of this method were tuned using the following setup. First, 3 linguists independently annotated whether or not each word from a random selection of 520 utterances from Voice 1 was emphasized. Then, \textit{WaveletBased} features were generated for the selected utterances using different values of the hyper-parameters. Each of these feature sets were quantized to 2 levels using k-means. The hyper-parameter set that gave the best match to the annotations in terms of F-score was chosen. The optimal hyper-parameters provided an F-score of 78\%.

80-dimensional Mel-spectrograms were computed from pre-emphasized speech using short-time Fourier transform with 25~ms frame length and 10~ms shift. All the emphasis control models were trained for 140k steps using 16 GPUs and a batch size of 512. A high-quality neural vocoder \cite{siva_2021_wavernn}, similar to the WaveRNN \cite{kalchbrenner2018efficient}, is trained for each speaker separately to convert the Mel-spectrograms to speech. The model consists of a single RNN layer with 512 hidden units, conditioned on Mel-spectrogram, followed by two dense layers with 256 hidden units each and a single soft-max sampling at the output. The model is trained with pre-emphasized speech, sampled at 24~kHz and $\mu$-law quantized to 8 bits for efficiency.

\subsection{Objective measures}
\label{sec:obj}

\begin{figure}[t]
    \centering
    \includegraphics[width=1.0\linewidth]{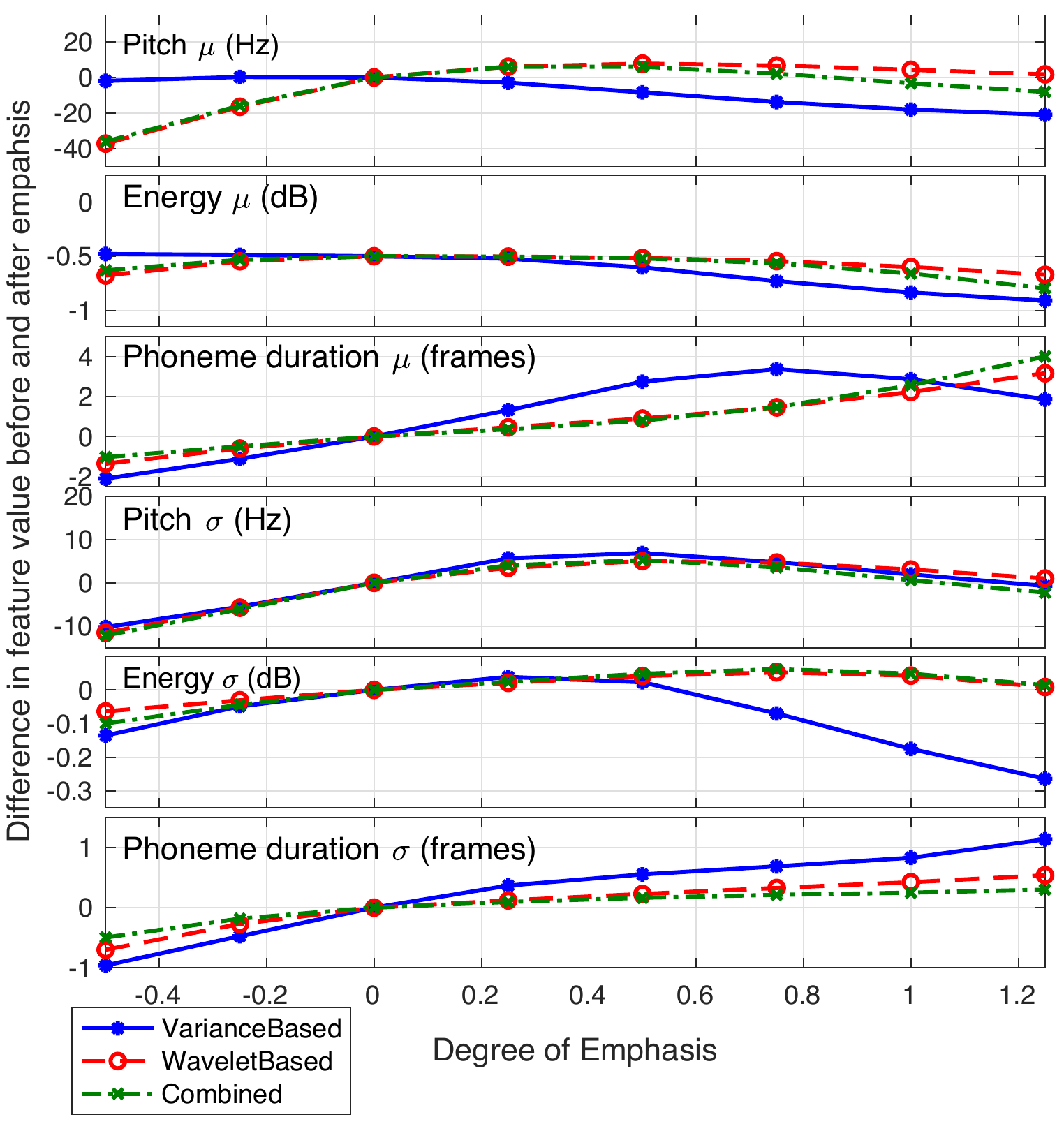}
    \vspace{-5.5mm}
    \caption{Difference in the mean ($\mu$) and standard deviation ($\sigma$) of pitch, energy, and phoneme duration after applying different degrees of emphasis, calculated over the phones of the emphasized word.}
    \vspace{-5mm}
    \label{fig:obj_comb}
\end{figure}

To analyze how the models with different emphasis features respond to varying degrees of emphasis control, we synthesized 192 test sentences for the {\it VarianceBased}, {\it WaveletBased}, and {\it Combined} models and each voice using a varying degrees of emphasis modification. This was done by adding different bias values to the predicted emphasis features for the phonemes of the emphasized word. Larger values of this bias should result in increased emphasis while negative values should correspond to de-emphasis. For the \textit{VarianceBased} method that has 2 emphasis features, both features were modified by the same amount. The emphasis features influence the pitch, energy, and phoneme duration of the models (see Fig.~\ref{fig:block_small}), and therefore, we measured the resulting changes in these three features before and after the changes in the emphasis features to assess the emphasis control. Fig.~\ref{fig:obj_comb} shows the change in the mean and standard deviations for pitch, energy, and phoneme duration after applying emphasis, averaged per phone in the emphasized words. The \textit{WaveletBased} and \textit{Combined} methods show similar behavior, relying mainly on pitch and phoneme duration means for emphasis modification. In contrast, with the \textit{VarianceBased} method, the mean pitch does not show much change for de-emphasis and is reduced when more emphasis is applied, and the standard deviation of phoneme duration is also increased to a greater degree. Energy does not seem to play a significant role in emphasis modification in any of the methods since the changes in energy are small and thus likely not perceivable.

\subsection{Subjective evaluation}
\label{sec:subj}

We performed three subjective evaluations assessing 1) perceived degree of emphasis, 2) perceived quality and 3) perceived preference with and without emphasis. We synthesized the 192 test utterances using the following 7 systems (for both speakers): the proposed systems with \textit{VarianceBased}, \textit{WaveletBased}, and \textit{Combined} features, before (3 systems) and after (3 systems) emphasis, as well as the \textit{Baseline} (1 system) model as shown in Fig.~\ref{fig:block_small} without any emphasis features. The emphasis levels were chosen so that they subjectively resulted to the same level of emphasis across the methods, which was 0.5 for \textit{VarianceBased} features and 0.75 for \textit{WaveletBased} and \textit{Combined} features, respectively. The tests were performed with American English native speakers using headphones.

\vspace{-1mm}
\subsubsection{Perceived degree of emphasis}

\begin{figure}[t]
    \centering
    \includegraphics[width=1.0\linewidth]{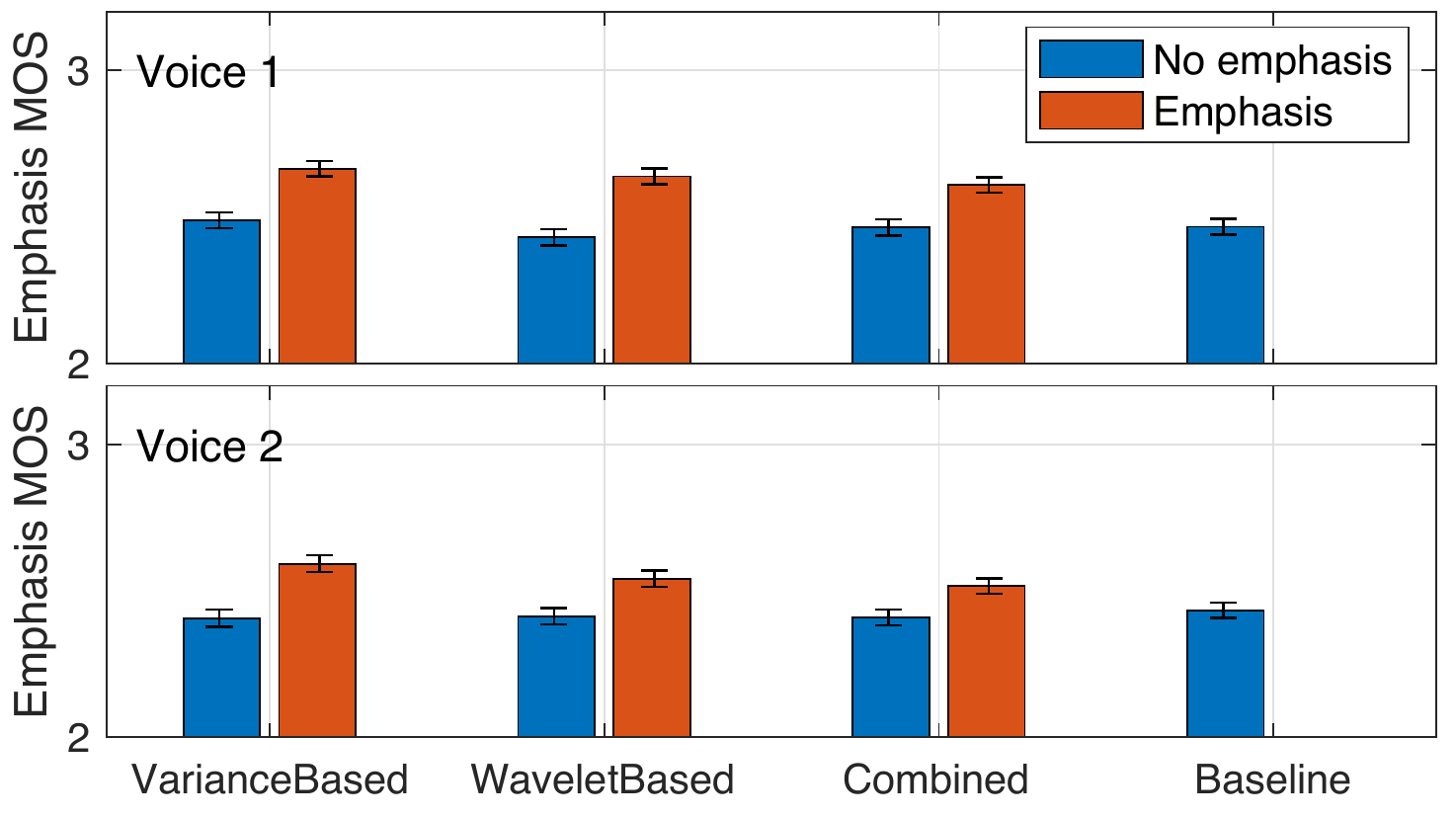}
    \vspace{-5mm}
    \caption{Results showing the perceived degree of emphasis: \textit{1-neutral, 2-slightly emphasized, 3-emphasized, and 4-strongly emphasized}. The y-axis is zoomed-in to highlight differences between the systems.}
    \vspace{-3mm}
    \label{fig:subj_emph}
\end{figure}

The test for perceived degree of emphasis was performed by 17 individual listeners, resulting in a total of 13,440 responses. Each of the raters were shown synthesized audio samples along with their corresponding text, highlighting (capitalized and with asterisk) the word that they should focus on. They were asked to rate the degree of emphasis of the highlighted word on a 4-point scale: 1-neutral, 2-slightly emphasized, 3-emphasized, and 4-strongly emphasized. Before the start of the test, the listeners were shown three examples and pointed out the reference emphasis level of the highlighted word. 

The results, depicted in Fig.~\ref{fig:subj_emph}, show that all the three methods achieve higher degree of emphasis after the emphasis modification. The three models achieve similar levels of increase in perceived emphasis. The \textit{VarianceBased} method for both voices and the \textit{WaveletBased} method for Voice 1 have a significantly higher level of emphasis after modification (measured using pairwise Student t-test with Bonferroni correction). Similarly, the level of perceived emphasis without emphasis modification is at the same level as the \textit{Baseline} model. Also, Voice 2 shows slightly lower increase in emphasis after modification compared to Voice 1.

\vspace{-1mm}
\subsubsection{Perceived quality}

\begin{figure}[t]
    \centering
    \includegraphics[width=1.0\linewidth]{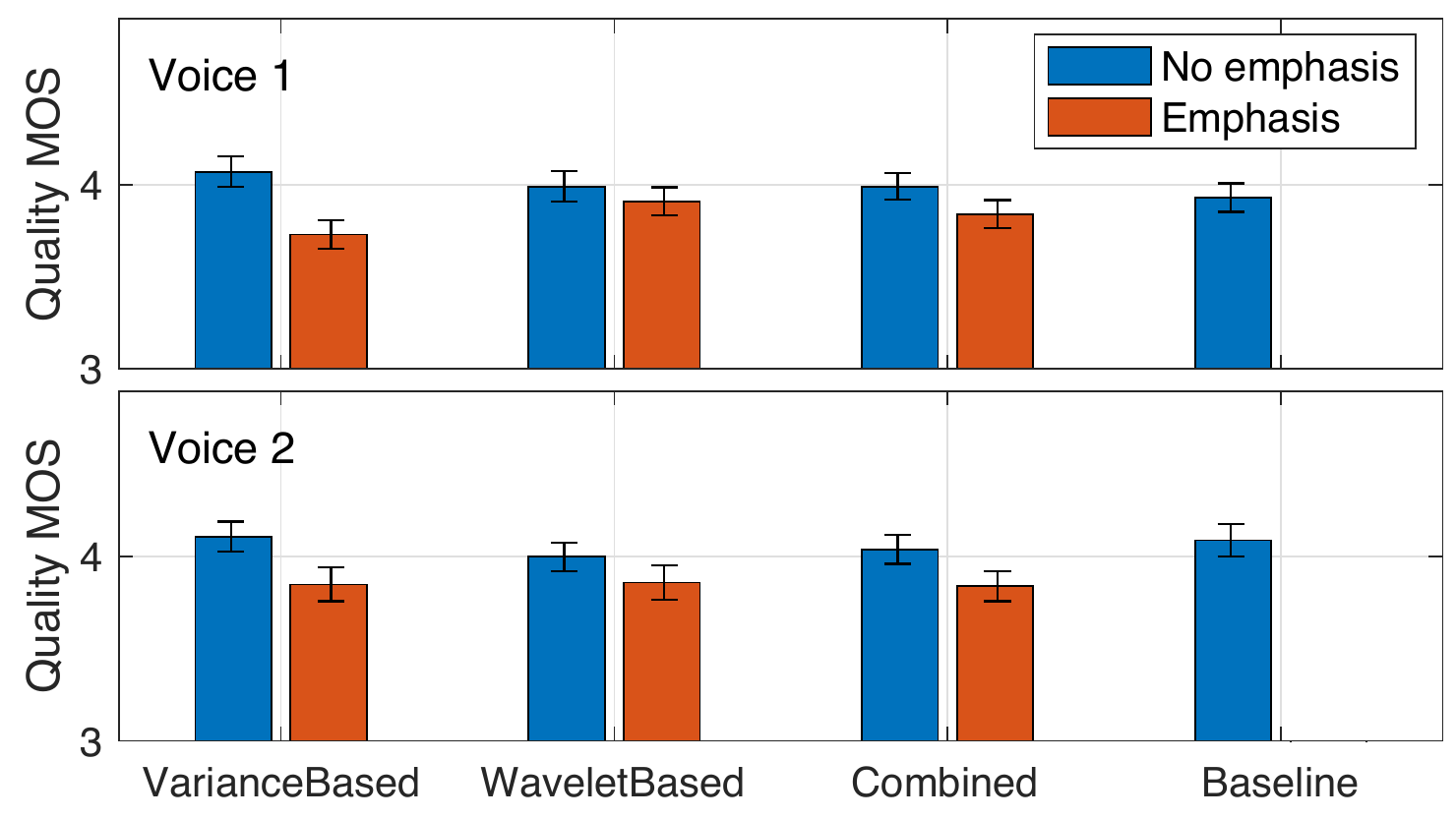}
    \vspace{-5mm}
    \caption{Results showing the  perceived quality: 1-bad, 2-poor, 3-fair, 4-good, 5-excellent. The y-axis is zoomed-in to highlight differences between the systems.}
    \vspace{-5mm}
    \label{fig:subj_mos}
\end{figure}

The test for perceived quality was performed by 27 individual listeners, resulting in a total of 7,890 responses. The raters were asked to rate the audio samples in terms of quality on a 5-point scale: 1-bad, 2-poor, 3-fair, 4-good, 5-excellent. The results, depicted in Fig.~\ref{fig:subj_mos}, show that all the models have a slightly reduced mean opinion score (MOS) after emphasis modification, with the \textit{VarianceBased}, and \textit{WaveletBased} methods showing the largest and lowest decrease, respectively. However, none of the quality MOS results were significantly different before and after emphasis modification (measured using pairwise Student t-test with Bonferroni correction). The quality MOS before emphasis modification for the three methods was found to be equal to the MOS of the \textit{Baseline} model.

\vspace{-1mm}
\subsubsection{Perceived preference}

\begin{figure}[t]
    \centering
    \vspace{-1mm}
    \includegraphics[width=1.0\linewidth]{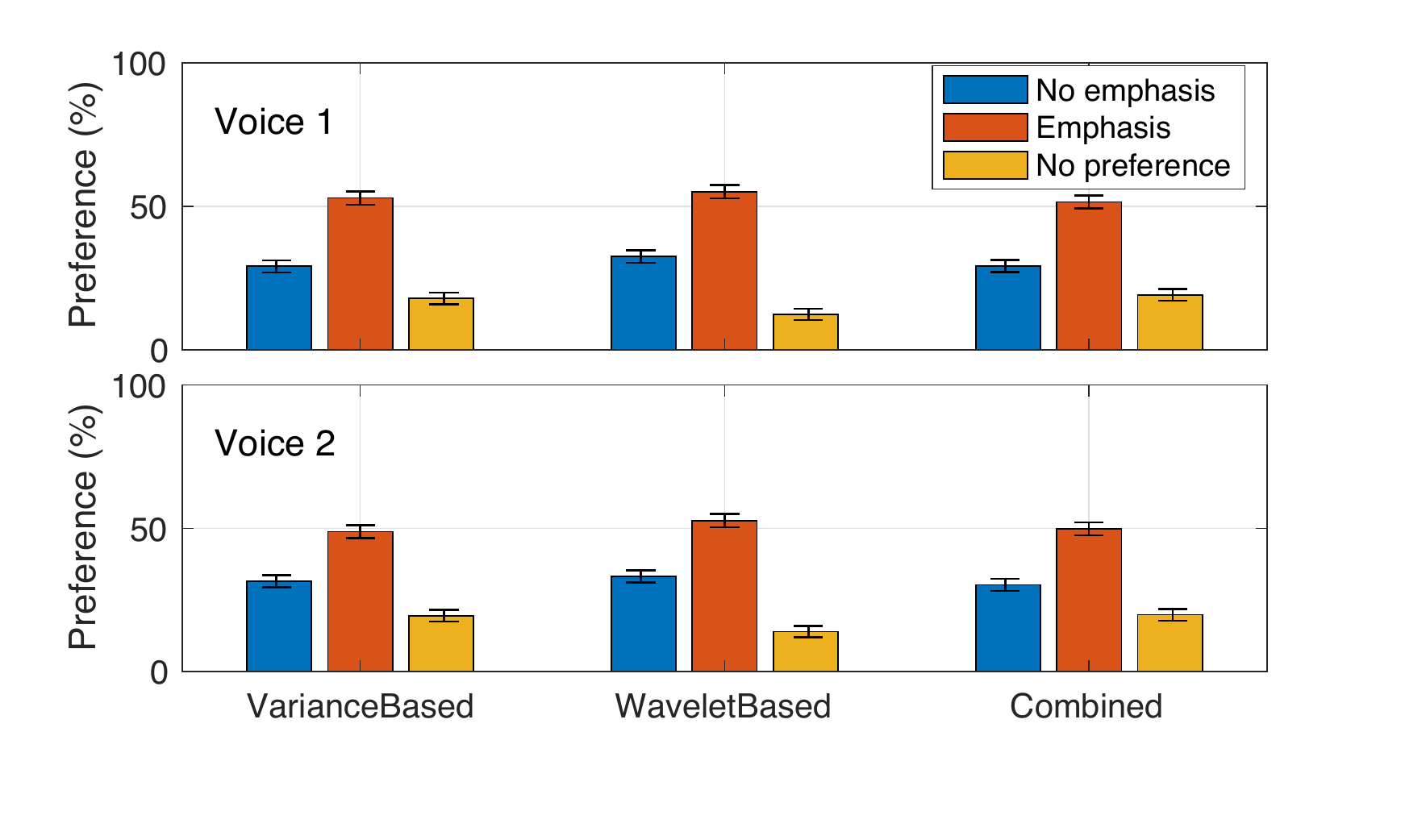}
    \vspace{-5mm}
    \caption{Results of the preference test.}
    \vspace{-4.5mm}
    \label{fig:ab}
\end{figure}

The test for perceived preference was performed by 29 individual listeners resulting in a total of 5,760 responses. The raters were shown a pair of utterances before and after emphasis modification from \textit{VarianceBased}, \textit{WaveletBased}, and \textit{Combined} models, and were asked to select which one they preferred (or neither). The results, depicted in Fig. \ref{fig:ab}, show that the utterances after emphasis modification were clearly preferred for all the three methods for both voices. All of the comparisons were significant as measured using the Pearson Chi-Squared test with Bonferroni correction.

\section{Discussion}
\label{sec:discussion}

The objective analysis shows that emphasis modification with different emphasis features result in slightly different changes in the output prosody. The \textit{WaveletBased} and \textit{Combined} methods mainly change the pitch of the word to be emphasized, whereas the \textit{VarianceBased} method has a much stronger effect on phoneme duration. Energy seems to be negligible for all the three methods, likely because the training data has limited variation in overall energy due to the audio leveling and compression used for preprocessing the audio. Moreover, increasing the emphasis features beyond what has been seen in the training data results mainly in the modification of phoneme duration.

Further analysis of the results revealed that some words, such as function words, short words, and words that are very unlikely to be emphasized in the training data, were not always properly emphasized while using the default level of emphasis. Therefore, the emphasis modification may need to be adjusted based on those factors to achieve proper emphasis.

Interestingly, the results of the perceived preference tests are in contrast to those of the perceived quality test. Although the exact reason for this is unclear, we hypothesize that in the perceived quality test, the raters focused more on the quality with respect to artefacts and naturalness of prosody without context. However, in the preference test, where both the emphasized and non-emphasized sentences were presented side by side, the raters may have simply preferred higher emphasis over lower emphasis due to the better suitability of emphasis in those specifically designed utterances, providing more distinct delivery of the information, and thus making them stand out and easier to listen to.

\section{Conclusions}
\label{sec:conclusion}

We presented a simple method for emphasis control for parallel neural TTS, and studied three different features for changing the degree of emphasis. The proposed approach enables an intuitive and low dimensional method for emphasis control. Subjective results show that the compared methods are able to provide a perceived increase in emphasis at the word level while maintaining good quality. Pairwise comparison tests also show an increased listener preference for the emphasized sentences.

\vfill\pagebreak

\bibliographystyle{IEEEtran}
\bibliography{mybib}

\end{document}